\documentclass[twocolumn,aps,showpacs,floatfix,superscriptaddress]{revtex4}
\usepackage{amsmath,amssymb,eucal,graphicx}
\begin{document}
\title{Smoothing a Rock by Chipping}
\author{P. L. Krapivsky and S.~Redner}
\email{paulk@bu.edu}
\email{redner@bu.edu}
\affiliation{Center for Polymer Studies and Department of Physics, Boston
University, Boston, Massachusetts 02215 USA}

\begin{abstract}
  We investigate an idealized model for the size reduction and smoothing of a
  polygonal rock due to repeated chipping at corners.  Each chip is
  sufficiently small so that only a single corner and a fraction of its two
  adjacent sides are cut from the object in a single chipping event.  After
  many chips have been cut away, the resulting shape of the rock is generally
  anisotropic, with facet lengths and corner angles distributed over a broad
  range.  Although a well-defined shape is quickly reached for each
  realization, there are large fluctuations between realizations.

\end{abstract}
\maketitle

\section{Introduction and Model}

This work is inspired by a recent experiment of Durian et al.\ \cite{DD} in
which they were interested in the ultimate shape of eroding rocks.  To
investigate this issue quantitatively, they studied the collisional erosion
of square clay particles due to their repeated impact with the walls of a
horizontally rotating plane enclosure.  In each such collision, chips break
off from the particle so that it gradually becomes rounder and smoother.  One
might naively expect the asymptotic particle shape to be a circle, as
protruding corners are more exposed and thus likelier to get rounded off by
this grinding process.

\begin{figure}[ht]
\hbox{\includegraphics*[width=0.4\textwidth]{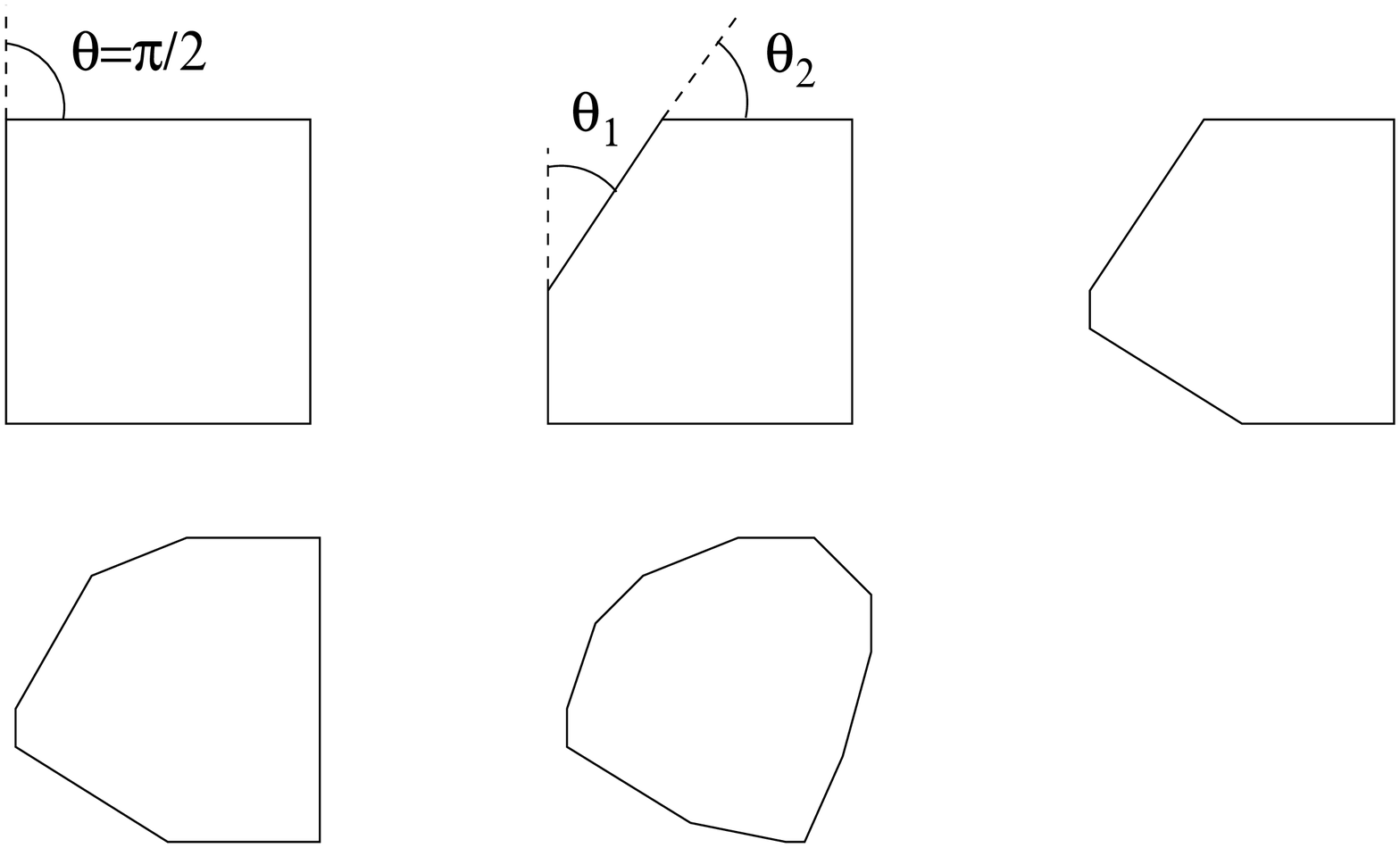}\hskip -0.6in\includegraphics*[width=0.15\textwidth]{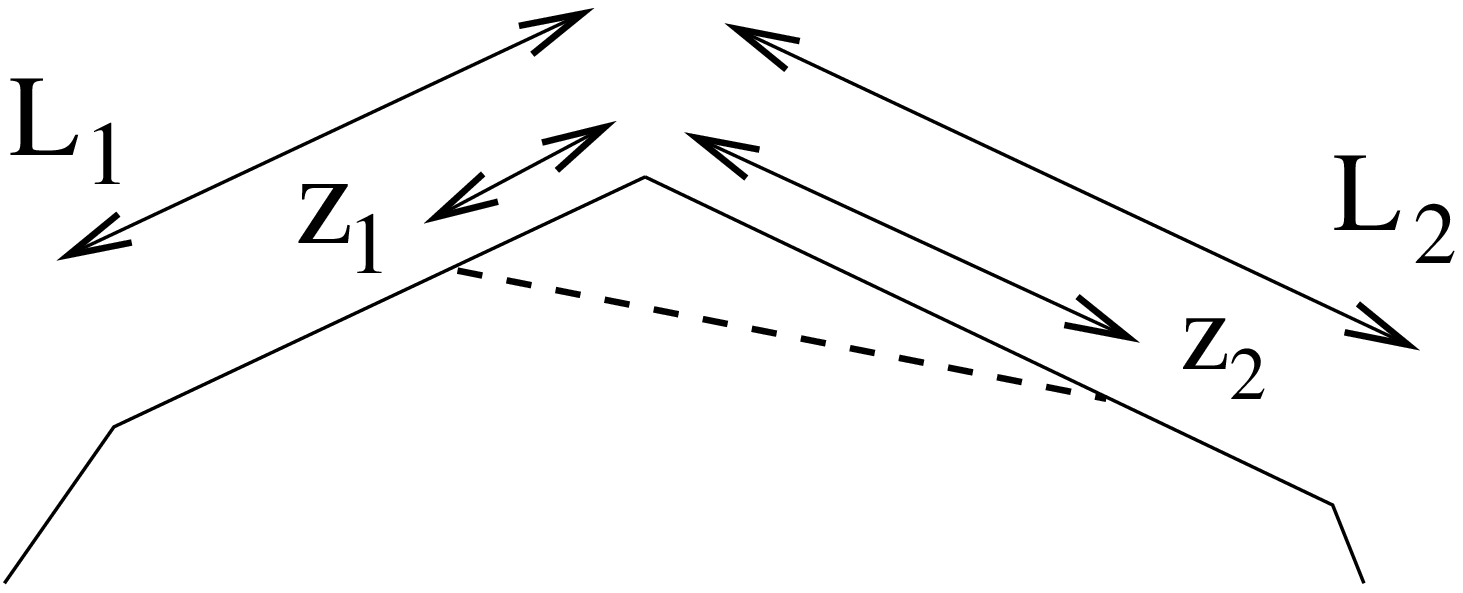}}

\caption{Schematic illustration of the cutting model after 1, 2, 3, and 8
  events.  A chip breaks from a randomly-selected corner with the sum of the
  new deflection angles, $\theta_1$ and $\theta_2$ equal to the deflection
  angle $\theta$ of the previous corner.  The geometry of a single chipping
  event is shown at the lower right. }
\label{model}
\end{figure}

\begin{figure}[ht]
\includegraphics*[height=1.27in,width=0.19\textwidth]{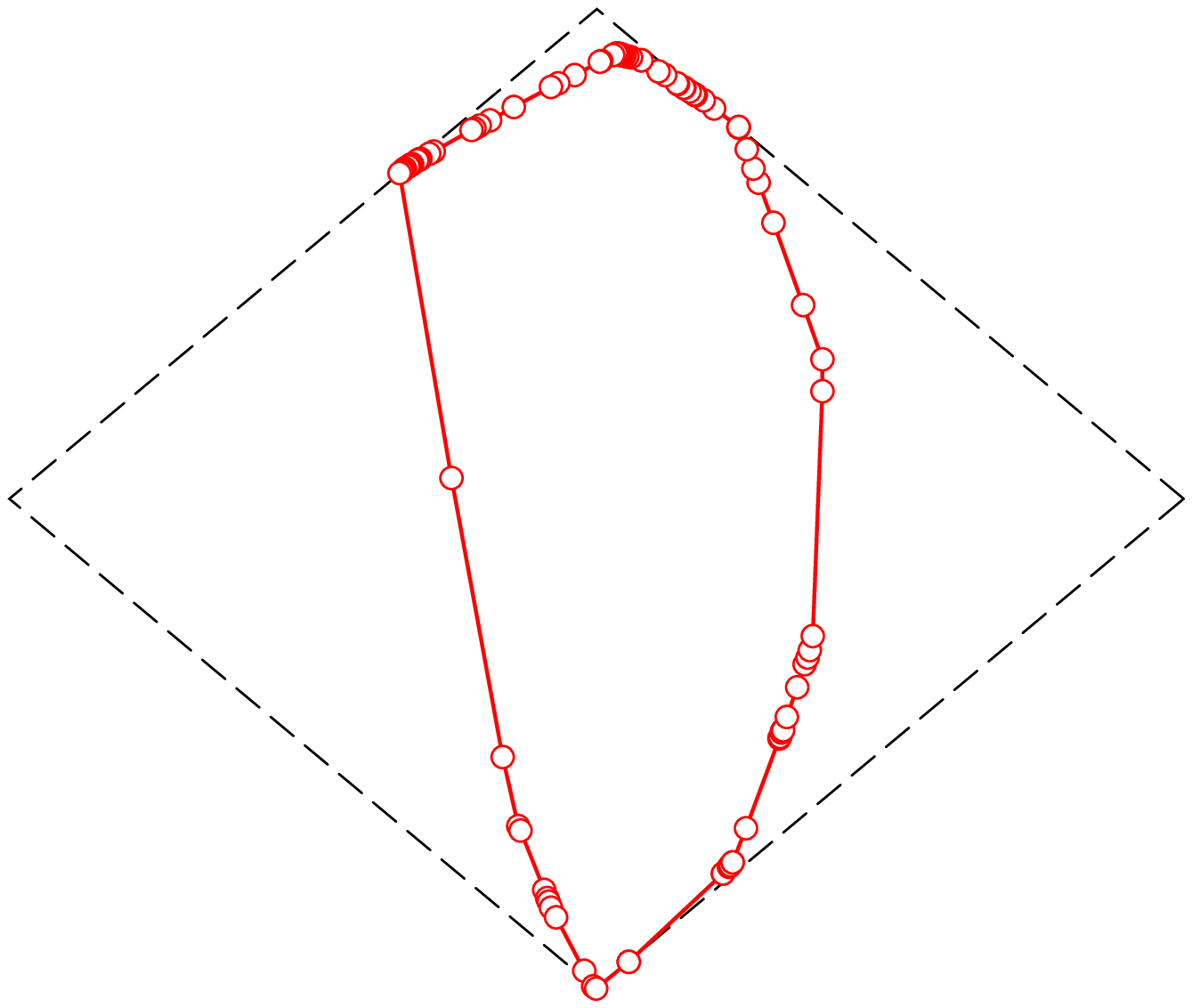}~~\includegraphics*[height=1.27in,width=0.19\textwidth]{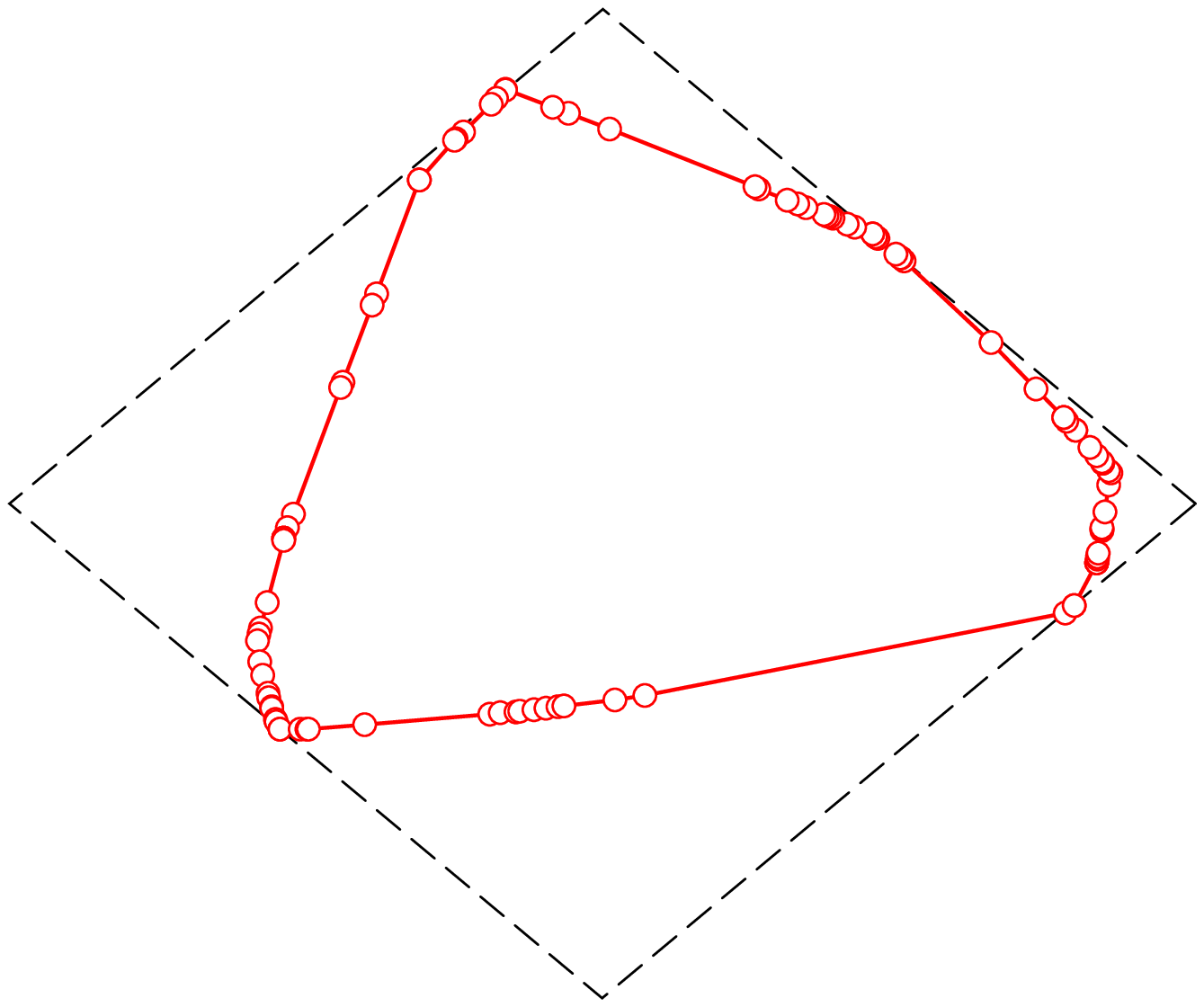}
\vskip 0.04in
\includegraphics*[height=1.27in,width=0.19\textwidth]{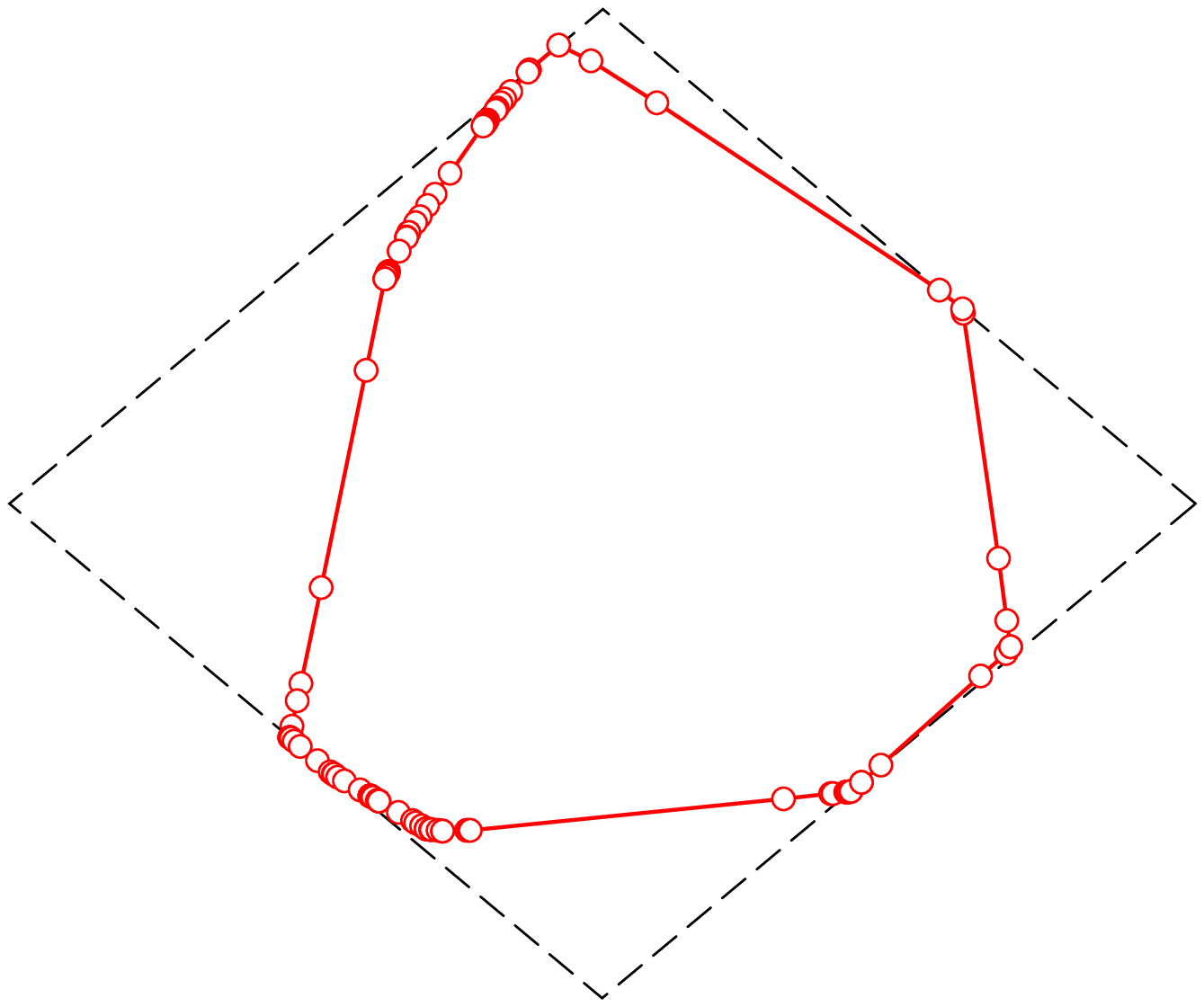}~~\includegraphics*[height=1.27in,width=0.19\textwidth]{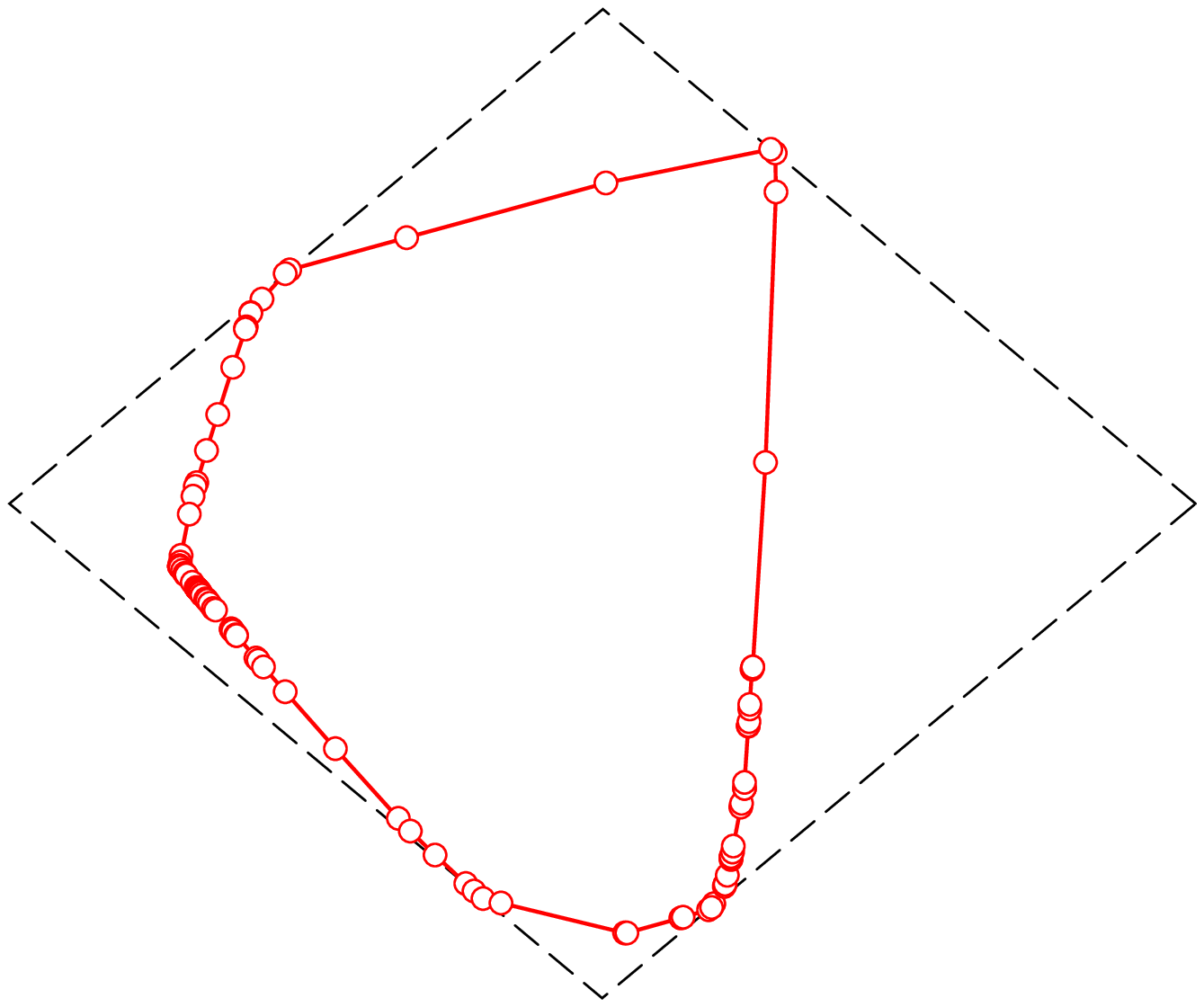}
\caption{(Color online) Four typical realizations of the cutting model for an
  initial unit square (dashed) after 100 corners (circles).}
\label{5r}
\end{figure}

A striking result from this experiment is that the asymptotic shape of the
eroding rock is not a circle \cite{DD}.  To describe this unexpected shape
evolution, Durian et al.\ also introduced a simple ``cutting'' model in which
the material exterior to a random chord on the object, whose length is
proportional to the square root of the remaining area, is removed in each
cutting event.  This step is repeated many times until asymptotic behavior is
reached.  Numerical simulations reproduced various aspects of the
experimental observations and confirmed that the asymptotic shape of the
particle is not circular \cite{DD}.

In this work, we investigate an idealized and analytically tractable version
of this cutting model.  A sequence of chipping events in our model is
schematically illustrated in Fig.~\ref{model}.  The rock is initially assumed
to be square.  In each chipping event, a piece of the rock is broken off at a
corner.  The deflection angles of the two newly-created corners sum to the
deflection angle of the original corner but are otherwise arbitrary.  The
sides $z_1$ and $z_2$ of a chip are smaller than the respective sides $L_1$
and $L_2$ of the corner itself (lower right in Fig.~\ref{model}), so that
only a single corner and a finite fraction of its two adjacent sides are
removed in each chipping event.  As the rock is chipped away, a non-trivial
shape is generated that is the focus of our interest.

\begin{figure}[ht]
\includegraphics*[width=0.25\textwidth]{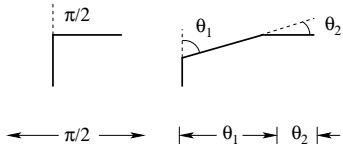}
\caption{Equivalence between angle evolution in chipping and fragmentation of
  a line segment.}
\label{asym}
\end{figure}

A convenient geometrical representation for the evolution of corner
deflection angles is to view the initial angle as a line segment of length
$\pi/2$ (Fig.~\ref{asym}).  Each chipping event then corresponds to picking
one segment at random and cutting it into two arbitrary size pieces.  This
connection to binary fragmentation allows us to make use of well-known
results for this latter problem \cite{ZM} to help understand geometrical
features of the object as its size is reduced by chipping.

The constraint that the chip breaks off only a single corner and a portion of
its two adjacent sides ensures that final particle size is strictly positive;
that is, the initial particle is not eroded to nothingness.  This fact can be
appreciated by examining Fig.~\ref{5r}.  By the definition of the model, a
short segment on each of the sides of the initial square must remain part of
the perimeter of the eroding particle in the long time limit.  This fact
ensures that a particle of non-zero size remains in the long-time limit.
Each of these four segments may be arbitrarily small and two such segments on
adjacent sides of the initial square may be arbitrarily close to each other.
In the exceptional case where short segments occur in pairs at opposite
diagonals of the initial square, the area can be arbitrarily small, but with
vanishing probability of the size being zero.  

In the next section, we investigate the evolution of the distribution of
facet angles by a master equation approach when each chipping event bisects
the corner angle.  We show that the resulting angle distribution has a
Poisson form in the variable $\ln t$, where $t$ is the current number of
corners.  In Sec.~\ref{sec-arb}, we study the evolution of the angle
distribution, as well as the actual shape of the object, for the general
situation where a chip divides a corner angle $\theta$ into two arbitrary
angles $\theta_1$ and $\theta_2$, with $\theta=\theta_1+ \theta_2$.  For this
process, we find: (i) the asymptotic shape of an eroding particle is not
round and (ii) after many chipping events, the particle is characterized by
large sample-to-sample fluctuations.  In Sec.~\ref{sec-ext}, we investigate
some natural extensions to more physical cutting rules and conclude that our
main qualitative results are robust with respect to these generalizations.
We close with a brief discussion in Sec.~\ref{sec-disc}.

\section{Angle Bisection}
\label{sec-angle}

\subsection{Master equation solution}
We first treat the special case in which a corner angle is always bisected in
each chipping event.  As a result of repeated chipping events a non-trivial
distribution of corner angles develops.  Starting with the four right-angle
corners of an initial square, after one chipping event, two angles of
magnitude $\pi/4$ are created, while three right-angle corners remain.  After
a second event, either two more $\pi/4$ angles are created by chipping a
right angle (this occurs with probability 3/5), or one $\pi/4$ angle is
replaced by two new $\pi/8$ angles (this occurs with probability 2/5).

To determine the angle distribution, it is convenient to introduce the
integer variable $k\equiv -\ln_2(2\theta/\pi)$.  The initial angle $\pi/2$
then corresponds to $k=0$, and each angle halving corresponds to $k$
increasing by 1.  We refer to a corner with deflection angle corresponding to
$k$ as a $k$-corner.  Let $n_k(t)$ be the {\em average} number of
$k$-corners.  Starting with a square, the initial condition is
$n_k(t=0)=4\delta_{k,0}$ and the number of corners at time $t$ is simply
$t+4$.  Then the change in $n_k(t)$ after one chipping event obeys the master
equation
\begin{equation}
\label{ME}
n_k(t+1)-n_k(t)=\frac{2}{t+4}\,n_{k-1}(t) -  \frac{1}{t+4}\, n_k(t)\,.
\end{equation}
The first term on the right side accounts for the gain of $k$-corners due to
the chipping of one of the $(k-1)$-corners at time $t$.  The probability of
this event is ${n_{k-1}}/(t+4)$, and each such event increases the number of
$k$-corners by 2.  Conversely, the second term accounts for the loss of
$k$-corners when one such corner is chipped.

The system of master equations is recursive, and they can be solved one by
one.  Since $n_{-1}\equiv 0$, the average number of the 0-corners
(right-angle corners) satisfies the closed equation
\begin{equation}
\label{n0}
n_0(t+1)= \frac{t+3}{t+4}\, n_0(t).
\end{equation}
Iterating this equation, the solution is
\begin{equation}
\label{n0-sol}
n_0(t)= \frac{12}{t+3}.
\end{equation}
The average number of 1-corners satisfies
\begin{equation}
\label{n1}
n_1(t+1)= \frac{t+3}{t+4}\, n_1(t)+\frac{2}{t+4}\,\frac{12}{t+3},
\end{equation}
with solution (subject to the initial condition $n_1(0)=0$)
\begin{equation}
\label{n1-sol}
n_1(t)= \frac{24}{t+3}\sum_{3\leq j\leq t+2}\frac{1}{j}.
\end{equation}
With the solution for $n_1$, the average number of 2-corners satisfies
\begin{equation}
\label{n2}
n_2(t+1)= \frac{t+3}{t+4}\, n_2(t)
+\frac{2}{t+4}\,\frac{24}{t+3}\sum_{3\leq j\leq t+2}\frac{1}{j},
\end{equation}
whose solution is
\begin{equation}
\label{n2-sol}
n_1(t)= \frac{48}{t+3}\sum_{3\leq i< j\leq t+2}\frac{1}{i\,j}.
\end{equation}

While these exact expression for $n_k$ become progressively unwieldy as $k$
increases, the asymptotic behavior follows easily by noticing that in the
$t\to\infty$ limit the master equation \eqref{ME} turns into a differential
equation
\begin{equation}
\label{DE}
\frac{d n_k}{d t} = -\frac{n_k}{t} +\frac{2}{t} \, n_{k-1}.
\end{equation}
These equations can also be solved straightforwardly in a sequential manner.
Using the fact that $n_0\sim {12}/{t}$ and rewriting \eqref{DE} as
\begin{equation}
\label{IF}
\frac{d (t n_k)}{dt} =\frac{2 (t n_{k-1})}{t},
\end{equation}
we then obtain the solution
\begin{equation}
\label{nk-final}
n_k(t) = \frac{12}{t}\, \frac{(2\ln t)^k}{k!}.
\end{equation}
Thus the logarithm of the angle is a Poisson distribution with $\langle
k\rangle = 2\ln t$, corresponding to $\langle\theta\rangle \propto e^{-t}$.

While the final result for $n_k$ is quite simple, we emphasize that
Eq.~\eqref{nk-final} refers to the number of $k$-corners averaged over all
possible realizations of the cutting model.  However, the actual number of
$k$-corners in a given realization, defined as $N_k$, may differ
substantially for $n_k\equiv\langle N_k\rangle$.  In the appendix, we
investigate some of the simplest features of the $N_k$ that illustrate their
strongly fluctuating nature. 

\subsection{Simulation results}
\label{subsec-sim}

We simulated the probabilistic rules underlying the recursion formula
\eqref{ME} to obtain the distribution $N_k$ for each realization.  This
numerical approach has the advantages of simplicity and efficiency, but with
the obvious disadvantage that the actual shape of the particle is not
accessible by this approach.

\begin{figure}[!ht]
\includegraphics*[width=0.26\textwidth]{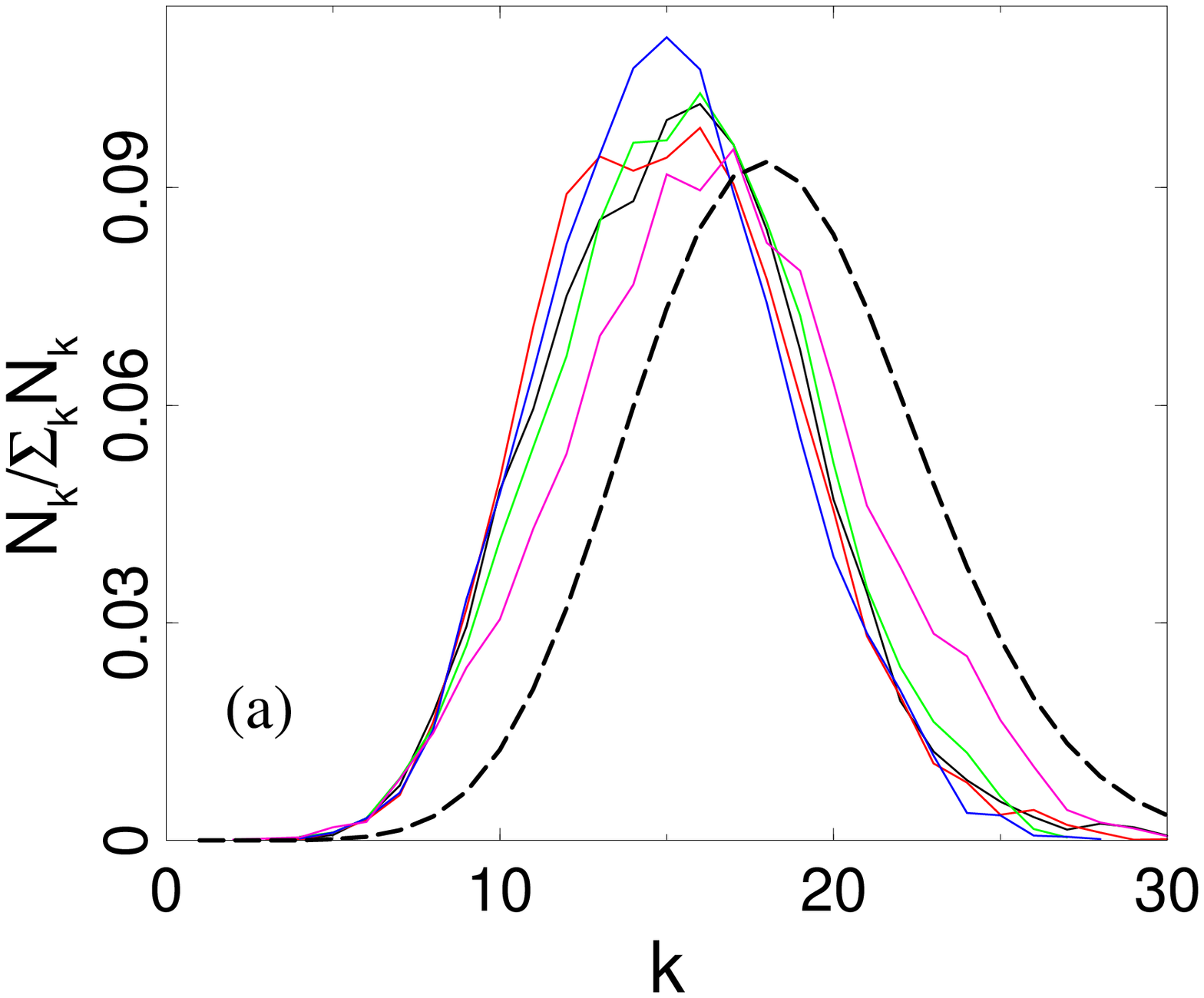}\includegraphics*[width=0.229\textwidth]{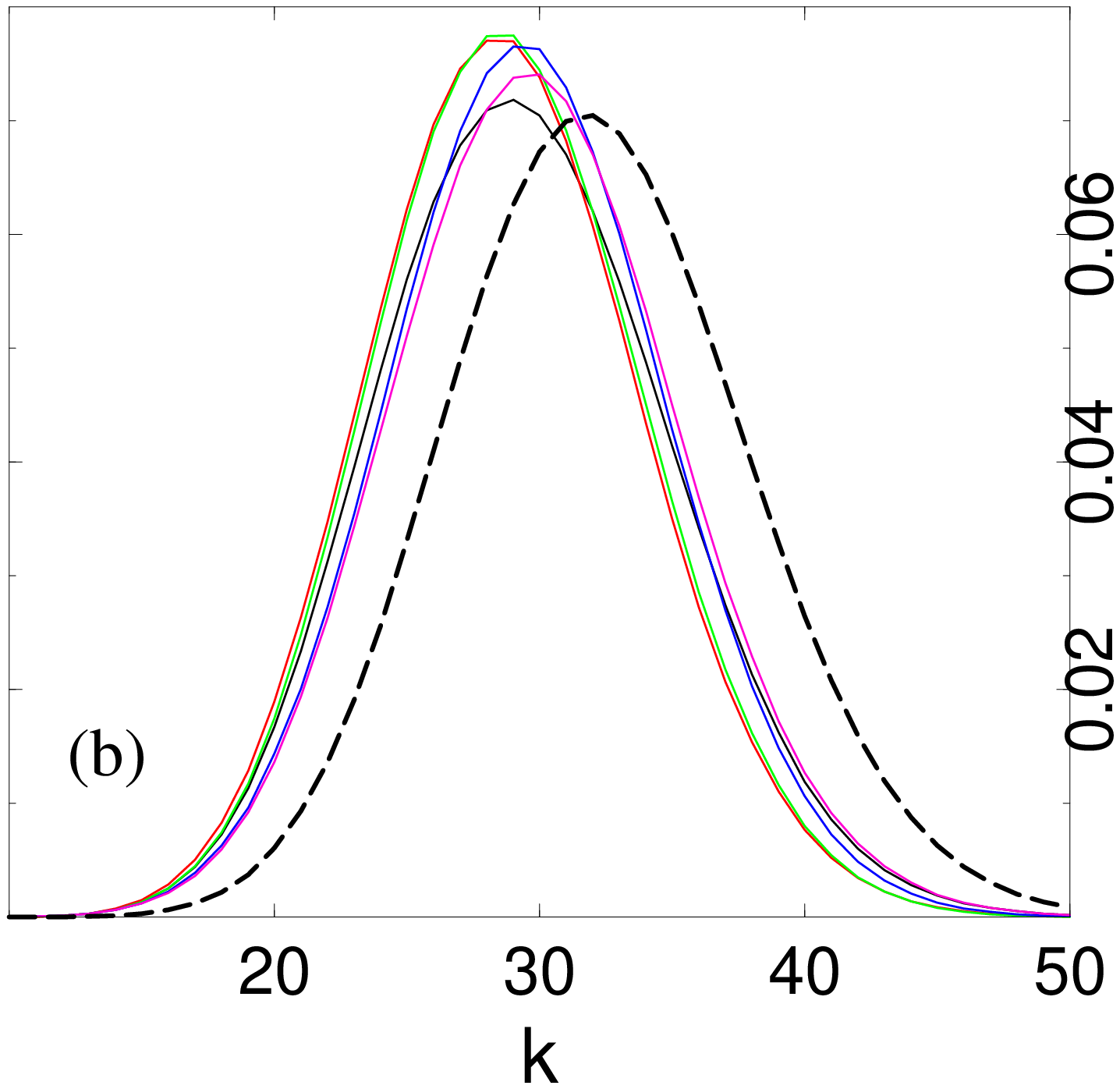}
\caption{(Color online) Normalized distributions of corner angles,
  $N_k/\sum_k N_k$, versus $k$ for 5 realizations of $10^4$ corners (a) and
  $10^7$ corners (b).  The heavy dashed curve is the properly normalized
  asymptotic expression from Eq.~\eqref{nk-final}.}
\label{sim}
\end{figure}

\begin{figure}[ht]
\includegraphics*[width=0.35\textwidth]{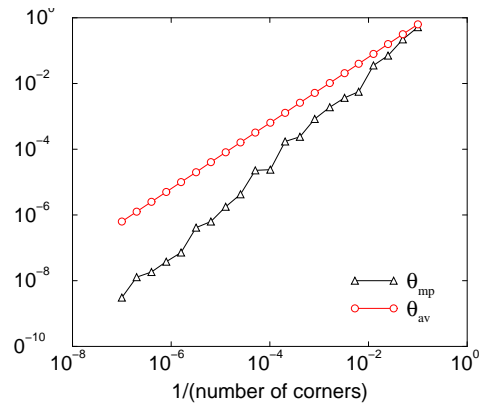}
\caption{(Color online) Average and most probable angles for a single
  realization of the cutting process as a function of the number of corners.}
\label{angles}
\end{figure}

The data show that the distribution of the logarithm of the angle (actually
$k=-\ln(2\theta/\pi)$) is close a Poisson form in $k$, as predicted by
Eq.~\eqref{nk-final}.  To compare the data with this analytic expression,
however, we need to properly normalize the latter.  Summing
Eq.~\eqref{nk-final} over all $k$, one obtains $\sum_k n_k=12t$ for the total
number of corners, whereas the exact result is $t+4$.  To correct for this
discrepancy, we therefore divide the expression in \eqref{nk-final} by 12 to
compare with the data in Fig.~\ref{sim}.  The data are in reasonable agreement
with the properly normalized analytic distribution, but it appears that one
would have to simulate the chipping process for an astronomical number of
corners to obtain good agreement between the data and the asymptotic
expression.

Another important property of the angle distribution is the large difference
between the average angle $\langle \theta\rangle$ and the most probable angle
$\theta_{\rm mp}$.  Because the distributions in Fig.~\ref{sim} are plotted
against $-\ln\theta$, it is clear that $\langle\theta\rangle$ is much larger
than $\theta_{\rm mp}$, which is located at the peak of the distribution as
shown in Fig.~\ref{angles}.

Finally, because the natural variable is $\ln \theta\,$ the actual angle
distribution is very broadly distributed.  Consequently, the asymptotic shape
of a particle as a result of this cutting process will not be circular.  A
related feature is that simulation results from different realizations are
visually quite different, as might be anticipated by the random
multiplicative process that underlies the chipping process.  We will discuss
this feature in more detail in the next section.

\section{Arbitrary Chipping Angles}
\label{sec-arb}

\subsection{The angle distribution}

We now study the general situation where a chipping event creates two unequal
angles.  To determine the resulting angle distribution, we make use of the
geometric connection between chipping and binary fragmentation.  As
illustrated in Fig.~\ref{asym}, a corner angle $\theta$ that becomes two
corners of angles $\theta_1$ and $\theta_2$ (with $\theta_1+\theta_2=
\theta$), corresponds to the length-conserving cutting of a segment of length
$\theta$ into two pieces of lengths $\theta_1$ and $\theta_2$.  The angle
distribution in chipping then corresponds to the length distribution in the
equivalent fragmentation process.

The length distribution may be solved using the techniques from the theory of
fragmentation \cite{ZM}.  For convenience, consider the scaled segment length
$x\equiv 2 \theta/\pi$.  Starting with a segment of scaled length $x=1$, the
master equation for the length distribution is
\begin{equation}
\label{frag}
\begin{split}
\frac{\partial c(x,t)}{\partial t} = - c(x,t)\int_0^x &F(y,x-y)\, dy\\
&+2\int_x^1 c(y,t) F(x,y-x)\,dy.
\end{split}
\end{equation}
Here $c(x,t)$ is the concentration of fragments of length $x$ at time $t$ and
$F(x,y)$ is the rate at which a fragment of size $x+y$ is cut two pieces of
sizes $x$ and $y$.  The first term on the right accounts for the loss of
fragments of size $x$ due to their fragmentation.  The total rate of these
events is $\int_0^xF(y,x-y)\, dy$.  The second term on the right accounts for
the creation of a fragment of size $y$ due to the breakup of a larger segment
of size $y$.

In many fragmentation processes \cite{ZM,CR}, the breakup rate $F(x,y)$ is a
homogeneous function of the form $F(x,y)=(x+y)^{\lambda-1}$.  That is, the
breakup rate of a cluster of size $x+y$ depends only on its size and not on
the size of the two daughter fragments.  To make a direct connection with
cutting, we require $\lambda=0$ so that the total breaking rate of a fragment
is independent of its size.  In this case, the master equation becomes
\begin{equation}
\label{frag-1}
\frac{\partial c(x,t)}{\partial t} = - c(x,t)+
2\int_x^1 c(y,t)\, \frac{dy}{y}.
\end{equation}
This master equation represents the generalization of \eqref{IF} to continuum
angles.

For the initial condition corresponding to a square, $c(x,t=0)=4\delta(x-1)$,
the distribution of unscaled fragment sizes at any later time is given by
\cite{ZM}
\begin{equation}
\label{frag-soln}
\begin{split}
  c(\theta,t)=\frac{8}{\pi}\sqrt{\frac{2t}{\ln(\pi/2\theta)}}& \, e^{-t}
  I_1\left(\sqrt{8t\ln(\pi/2\theta)}\right)\\
  &+\frac{8}{\pi} e^{-t}\delta\Big(\theta-\frac{\pi}{2}\Big),
\end{split}
\end{equation}
where $I_1$ is a modified Bessel function of order 1.  The second term on the
right-hand side corresponds to the probability that there have been no
fragmentation events up to time $t$, while the first term on the right gives
the scaling part of the fragment size distribution.

From the variable combination in this first term, we find that the
characteristic angle has the time dependence $\theta \sim e^{-t}$.
Furthermore, from the asymptotic form of the Bessel function \cite{AS}, the
distribution of the logarithmic angle has a stretched exponential tail
\begin{eqnarray*}
c(x,t)\sim e^{\sqrt{-t\ln\theta}},
\end{eqnarray*}
with $-\ln\theta$ being the natural variable of the system.  As in the case
of symmetric chipping (angle bisection), the general chipping process leads
to a broad distribution of angles.  This result again suggests that the
asymptotic shape of the particle is not circular.

\subsection{Shape Evolution}
\label{subsec-shape}

\noindent{\it Area and Perimeter Distributions.}
Two fundamental characteristics of an object's shape are its area and its
perimeter.  Starting with a unit-area square, the resulting area and
perimeter distributions become smooth, sharply peaked about their average
values, and visually independent of the number of corners $N$ when $N\agt
20$.  The support of the area distribution is $[0,1]$, with a peak near 0.67.
Similarly, the support of the perimeter distribution is $[2\sqrt{2},4]$ and
the peak occurs at approximately 3.3.

\begin{figure}[ht]
\includegraphics*[width=0.35\textwidth]{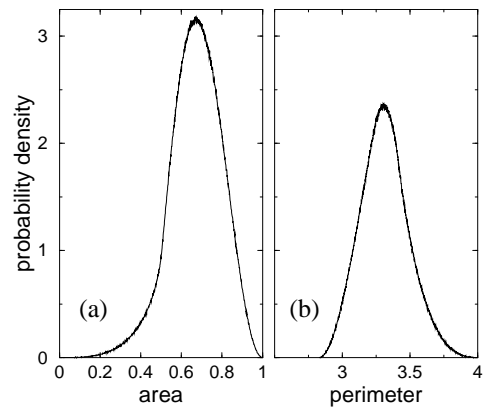}
\caption{The distribution of (a) area and (b) perimeter after 50
  corners for $10^7$ realizations.}
\label{area-perim-dist}
\end{figure}

An amusing unexplained feature is that a careful examination of the data
reveals that the first derivatives of both distributions are actually
discontinuous---the area distribution at area equal to 1/2 and perimeter
distribution at scaled perimeter $p_{\rm scaled}\equiv ({p-p_{\rm
    min}})/({p_{\rm max}-p_{\rm min}})$ also at $p_{\rm scaled}=1/2$.

\medskip\noindent{\it Asymmetry and Fluctuations}.  After a particle has
approximately 50 corners, a given realization is visually close to its
asymptotic shape.  As illustrated in Fig.~\ref{5r}, large fluctuations
between different realizations arise, so that the shape of a single
realization has little connection to the average shape.

\begin{figure}[ht]
\includegraphics*[width=0.35\textwidth]{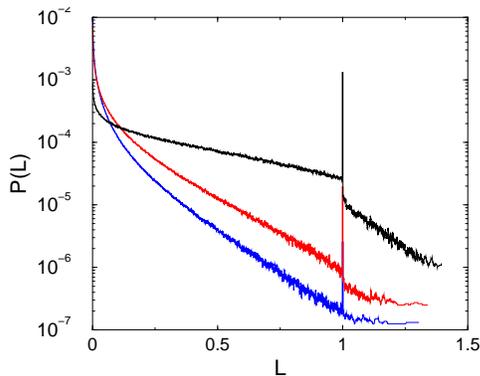}
\caption{(Color online) Probability distribution of facet lengths $P(L)$
  versus $L$ for $10^5$ realizations of 10, 40, and 80 corners.  The steeper
  curve corresponds to larger $N$. }
\label{facet-dist}
\end{figure}

Each individual interface typically consists of a few longer facets that are
punctuated by regions with many short facets, with a consequent large change
in the local tangent.  To illustrate this punctuated interface, we show the
facet length distribution for $10^5$ realizations with $N=10, 40$, and 80
corners (Fig.~\ref{facet-dist}).  The spike at $L=1$ corresponds to the
initial unit-length facets that remain unchipped.  The tail for $L>1$
corresponds to an initial cut that is sufficiently close to the main diagonal
of the initial square so that the facet length can be greater than 1---in
fact, the maximal facet length is $\sqrt{2}$.  This large-$L$ tail is
distinct from the rest of the distribution when the number of corners is
small.  As the number of corners increases, the number of short facets
correspondingly increases, and there is a huge buildup of the small-length
tail of the facet length distribution.

\begin{figure}[ht]
\includegraphics*[width=0.35\textwidth]{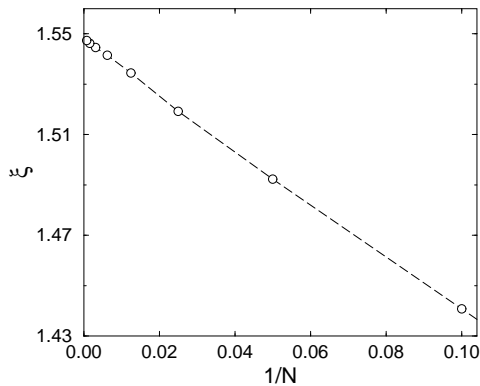}
\caption{Asymmetry ratio $\xi(N)\equiv \sqrt{\langle R_+^2(N)\rangle}
  /\sqrt{\langle R_-^2(N)\rangle}$ versus $1/N$.  Each data point is based on
  $10^6$ realizations, with $N=10, 20,\ldots, 1280$. }
\label{asymm}
\end{figure}

Finally, we study how the asymmetry of the particle evolves during the
cutting process.  The proper measure of asymmetry is through the moment of
inertia tensor of an object.  For the cutting model, this leads to a
cumbersome calculation when the number of corners is large.  We therefore
adopt a simpler approach that should reveal the same type of information
as the inertia tensor.  After an initial square has been reduced to an object
with a specified number of corners $N$, we determine the $x$- and
$y$-coordinates of each corner about the center of the initial square and
then compute the mean-square displacements of the $x$- and $y$-coordinates of
all the corners
\begin{eqnarray*}
X^2(N)=\frac{1}{N} \sum_{i=1}^N x_i^2\qquad Y^2(N)=\frac{1}{N} \sum_{i=1}^N y_i^2
\end{eqnarray*}
in each realization.  For each realization we then define the larger and the
smaller of these two mean-square displacements,
\begin{eqnarray*}
R_+^2(N)&= {\rm max}(X^2(N),Y^2(N))\\
 R_-^2(N)&= {\rm min}(X^2(N),Y^2(N)),
\end{eqnarray*}
and then average these maximal and minimal mean-square radii over many
realizations.  Finally, we quantify the asymmetry by the dimensionless ratio
\begin{eqnarray*}
\xi(N)\equiv
\sqrt{\langle R_+^2(N)\rangle}/\sqrt{\langle R_-^2(N)\rangle}.
\end{eqnarray*}
Thus, for example, a rhombus in which the two corners in the $x$-direction
are a unit distance from the origin while the other two corners are a
distance $1+\epsilon$, $\xi=(1+\epsilon)$.  As a function of $1/N$, the data
for $\xi(n)$ are quite linear for $N$ between 10 and 1280.  Extrapolating the
data of Fig.~\ref{asymm} to $1/N\to 0$, we infer the value of
$\xi(N\to\infty)\approx 1.548$, with a subjective uncertainty of 0.001.

\section{Extensions of the Model}
\label{sec-ext}

An unrealistic feature of our cutting model is that each corner has the same
probability of being chipped.  As a consequence, corners tend to congregate,
as seen in Fig.~\ref{5r}.  In the equivalent fragmentation process,
equiprobable corner chipping corresponds to an overall breakup rate for a
given segment that is {\it independent\/} of its length.  This
length-independent breakup rate in fragmentation demarcates the boundary
between scaling solutions, when the breakup rate grows with segment length,
and ``shattering'' solutions \cite{ZM,CR}, when the breakup rate decreases
with segment length.  The shattering solution is characterized by a finite
fraction of the system being transformed into a dust of zero-length particles
that contain a finite fraction of the initial length.  This singularity is
parallel to the gelation transition in irreversible aggregation.
 
Thus a natural question is whether different behavior arises in the
physically more realistic situation in which larger protrusions are likelier
to be chipped.  In the language of the equivalent fragmentation process, we
should study break-up rates with a positive homogeneity index---namely,
larger fragments are more likely to break.  To study the role of a positive
homogeneity index, we considered the extreme situation in which only the most
susceptible corner breaks in a chipping event.  We thus investigated the
following three extremal dynamics rules:

\begin{enumerate}
\itemsep -2pt
\item Chip the corner furthest away from the origin.
\item Chip one of the corners on the longest facet.
\item Chip the corner with the largest deflection angle.
\end{enumerate}

Each of these chipping rules focuses on some aspect of the most prominent
non-smooth regions of the object.  Qualitatively, we find that these three
rules all lead to a non-circular asymptotic shape of the object.  The reason
for this non-circularity ultimately stems from the strong role played by the
first few chipping events.  The size of each chip can, in principle, range
from zero to its maximum attainable size (see Fig.~\ref{model}).  If one of
these early chips is close to its maximal size, this chip leaves an imprint
on the object that persists in the long-time limit.  This property is
different than that of curvature-driven interface evolution, in which the
amount by which a curved region of the interface moves is strictly
proportional to the local curvature \cite{curvature}.  

\begin{figure}[ht]
\includegraphics*[width=0.25\textwidth]{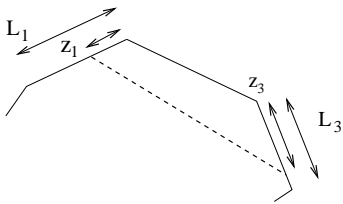}
\caption{Geometry of a two-corner chipping event.}
\label{2c}
\end{figure}

\begin{figure}[ht]
\includegraphics*[width=0.365\textwidth]{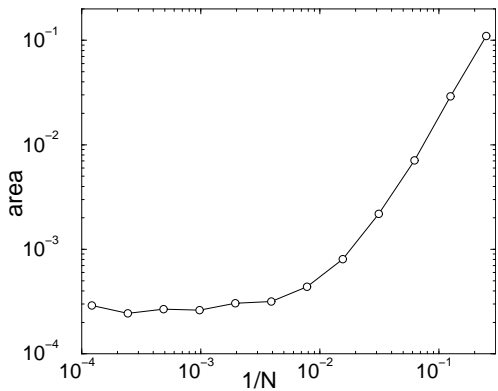}
\caption{Average area of a unit square after $N$ cuts versus $1/N$ on a
  double logarithmic scale when the probability of 2-corner chipping
  $p_2=0.9$.}
\label{area-0.9}
\end{figure}

Another natural concern about the applicability of the cutting model is the
restriction to breaking only a single corner and portions of its adjacent
sides in a single chipping event.  Indeed this rule ensures that the final
size of the particle remains non zero as mentioned in the introduction.  To
test the robustness of the cutting model results to the possibility that more
than one corner can be chipped away, we studied the situation in which a chip
could encompass two corners, as illustrated in Fig.~\ref{2c}.  Specifically,
we pick a corner at random; with probability $p_2$, the chip includes both
this corner and its nearest neighbor.  With this rule, the restriction that a
small segment of the initial square must remain as part of the boundary of
the eroding particle no longer applies.  Thus it is not obvious {\it a
  priori} that the size of the object will remain non-zero in the long-time
limit.  Nevertheless, this more generous two-corner chipping rule still leads
to a non-zero particle size, as long as the probability for single-corner
chips is non-zero. Initially, the area decreases rapidly as the number of
cuts increases.  However, As the number of corners becomes appreciable, later
cuts remove only a tiny fraction of the particle so that the area eventually
saturates to a non-zero value (Fig.~\ref{area-0.9}).

\section{Discussion}
\label{sec-disc}

We studied the geometric properties of an idealized model for the erosion of
a two-dimensional rock by repeated chipping of small pieces.  A chipping
event is defined by cutting a small piece from the rock in which a single
corner and part of its two adjacent sides are removed.  In our model, each
corner has the same probability of being chipped.  A two basic outcome of
this cutting model is that there is shape asymmetry in the long-time limit.
Thus the asymptotic outcome after many chipping events is not a circle, as
was initially observed in the experiments and the simulations of Durian et
al.\ \cite{DD}.  Another important feature is that there are large shape
fluctuations between realizations so that the outcome of a single event is
not representative of the average behavior.

We determined the evolution of the distribution of angles from its governing
master equation.  For the case of angle bisection in each chipping event, we
found a broad and asymptotically Poissonian distribution of angles in the
variable in $\ln t$, where $t$ is the number of chipping events.  This
behavior appears to be idiosyncratic to the case of angle bisection.  In the
more realistic case where a chipping event divides an initial angle into two
arbitrary angles (with conservation of the total angle), the angle
distribution has a different behavior that is immediately obtained by the
exact correspondence between the distribution of angles in the cutting model
and the distribution of fragments sizes in the binary fragmentation of a line
segment \cite{ZM,CR}.

Finally, it is worth mentioning that because of large sample-to-sample size
and shape fluctuations after a given number of chipping events, the cutting
model does not give a unique limiting shape.  This behavior is in contrast to
the class of interface models where the asymptotic shape of a single
realization of interface converges to a unique limiting shape.  Two famous
such examples are the strictly convex interface between the origin and
$(x\gg0,y\gg0)$ \cite{segment} and the interface that is generated by the
partition of the integers \cite{partition}.  It may be worthwhile to explore
variants of the cutting model that lead to a unique limiting shape to take
advantage of the highly-developed analysis methods available for this type of
interface evolution process.  It should also be of interest to extend our study
to the more realistic case of three dimensions, where there is also a
highly-developed mathematical literature on limiting shapes \cite{3d}.

\acknowledgments{ We thank Doug Durian for helpful correspondence, Michael
  Kearney for providing manuscript corrections, and financial support from
  NSF grants CHE0532969 (PLK) and DMR0535503 (SR).}

\appendix*

\section{Fluctuations in $n_k$}

In this appendix we investigate the statistical properties of the
distribution of $k$-corners in greater detail.  We show that the actual value
of the number of $k$-corners in a given realization, $N_k$, is generally
quite different than the average number of $k$-corners, $n_k$.  This lack of
self averaging can be seen by studying the probability distribution of the
random quantities $N_k(t)$ across all realization.  While the full
calculation is a tedious endeavor, it is fairly simple to obtain the number
of right-angle corners $N_0(t)$.  The master equation for $N_0$ is simpler
than that for all the other $N_k$ with $k>0$, because $N_0$ can never
increase in a single chipping event.

Starting with a square that has four right-angle corners, the number of such
corners is a deterministic quantity when $t=0$ and $t=1$,
\begin{equation*}
N_0(0)=4, \quad N_0(1)=3,
\end{equation*}
while for $t>1$  the number of right-angle corners is a random quantity.  Let
 \begin{equation*}
\Pi_j(t)\equiv {\rm Prob}\{N_0(t)=j\},
\end{equation*}
and let us compute $\Pi_j(t)$ for $t>1$.  

We begin with $\Pi_3(t)$.  To have three right-angle corners, each chipping
event must not act on any of these corners.  Consequently, the probability to
have three right-angle corners satisfies the recurrence
\begin{equation}
\label{Pi3-rec}
\Pi_3(t+1)=\frac{t+1}{t+4}\,\Pi_3(t).
\end{equation}
Solving this recurrence subject to the initial condition $\Pi_3(1)=1$ we obtain
\begin{equation}
\label{Pi3}
\Pi_3(t)=\frac{24}{(t+1)(t+2)(t+3)}.
\end{equation}
By similar reasoning, the recurrence for the probability to have two
right-angle corners is
\begin{equation}
\label{Pi2-rec}
\Pi_2(t+1)=\frac{t+2}{t+4}\,\Pi_2(t)+\frac{3}{t+4}\,\Pi_3(t),
\end{equation}
whose solution is
\begin{equation}
\label{Pi2}
\Pi_2(t)=\frac{36(t-1)}{(t+1)(t+2)(t+3)}.
\end{equation}
The probability to have a single right-angle corner satisfies the recurrence
\begin{equation}
\label{Pi1-rec}
\Pi_1(t+1)=\frac{t+3}{t+4}\,\Pi_1(t)+\frac{2}{t+4}\,\Pi_2(t),
\end{equation}
from which
\begin{equation}
\label{Pi1}
\Pi_1(t)=\frac{12(t-1)(t-2)}{(t+1)(t+2)(t+3)}.
\end{equation}
Finally the probability that there are no right-angle corners can be found by
solving the appropriate recurrence formula
\begin{equation}
\label{Pi0-rec}
\Pi_0(t+1)=\Pi_0(t)+\frac{1}{t+4}\,\Pi_1(t),
\end{equation}
or from the normalization $\Pi_0+\Pi_1+\Pi_2+\Pi_3=1$.  In either case, we
obtain
\begin{equation}
\label{Pi0}
\Pi_0(t)=\frac{(t-1)(t-2)(t-3)}{(t+1)(t+2)(t+3)}
\end{equation}

As a useful consistency check one can compute
\begin{equation*}
n_0= \langle N_0\rangle=\sum j\Pi_j\,
\end{equation*}
and recover the result $n_0=\frac{12}{t+3}$ given in Eq.~\eqref{n0-sol}.


\begin{thebibliography}{99}

\bibitem{DD} D.J. Durian, H. Bideaud, P. Duringer, A. Schroder, F. Thalmann,
  and C.M. Marques, Phys.\ Rev.\ Lett.\ {\bf 97}, 028001 (2006); D.J. Durian,
  H. Bideaud, P. Duringer, A. Schroder, and C.M. Marques, cond-mat/0607122.

\bibitem{ZM} R. M. Ziff and E. D. McGrady, J. Phys.\ A {\bf 18}, 3027 (1985);
  R. M. Ziff, J. Phys.\ A {\bf 25}, 2569 (1992).

\bibitem{CR} Z. Cheng and S. Redner, Phys.\ Rev.\ Lett.\ {\bf 60}, 2450
  (1988); J. Phys.\ A {\bf 23}, 1233 (1990).

\bibitem{AS} M. Abramowitz and I. A. Stegun {\it Handbook of Mathematical
    Functions} (Dover, New York, 1972).

\bibitem{curvature} W. W. Mullins, J. Appl.\ Phys.\ {\bf 27}, 900 (1956); G.
  Huisken, J. Diff.\ Geom.\ {\bf 20}, 237 (1984); M. E. Gage and R. S.
  Hamilton, J. Differential Geom.\ {\bf 23}, 69 (1986); M. A. Grayson, J.
  Diff.\ Geom.\ {\bf 26}, 285 (1987); D. L. Chopp and J. A. Sethian,
  Exp.\ Math.\ {\bf 2}, 235 (1993); D. L. Chopp, Exp.\ Math.\ {\bf 3}, 1
  (1994).

\bibitem{segment}
  A.~Vershik, 
  Func.\ Anal.\ Appl.\ {\bf 28}, 13 (1994);
  Ya.~G.~Sinai, 
  Func.\ Anal.\ Appl.\ {\bf 28}, 108 (1994).

\bibitem{partition} A.~Vershik, Func.\ Anal.\ Appl.\ {\bf 30}, 90 (1996);
  S.~Shlosman, J. Math.\ Phys.\ {\bf 41}, 1364 (2000); P.~L.~Krapivsky,
  S.~Redner, and J.~Tailleur, Phys.\ Rev.\ E {\bf 69}, 026125 (2004); R.
  Rajesh and D. Dhar, Phys.\ Rev.\ E {\bf 71}, 016130 (2005).

\bibitem{3d} A.~Okounkov and N.~Reshetikhin, J. Am.\ Math.\ Soc.\ {\bf 16}, 581  (2003);
A.~Okounkov and N.~Reshetikhin, Commun.\ Math.\ Phys.\  {\bf 269}, 571 (2007);
R.~Kenyon and A.~Okounkov, 
math-ph/0507007.
\end{thebibliography}
\end{document}